\documentclass[12pt]{article}
\usepackage[dvipdfmx]{color}
\usepackage{graphicx}
\usepackage{xcolor}
  \usepackage{amsthm,amsfonts}
  \usepackage{amsmath}
\usepackage{mathabx}
\usepackage{slashed}
\usepackage{ulem}
\usepackage{hyperref}
\usepackage{cleveref}
\usepackage{cite}

\newcommand{\bea}   {\begin{eqnarray}}
\newcommand{\eea}   {\end{eqnarray}}

\begin{document}
\renewcommand{\thefootnote}{\fnsymbol{footnote}}

\thispagestyle{empty}

\title{ A colored look at anyons and parastatistics}
\author{ Francesco Toppan\thanks{{E-mail: {\it toppan@cbpf.br}}}
\\
}
\maketitle

{\centerline{
{\it CBPF, Rua Dr. Xavier Sigaud 150, Urca,}}\centerline{\it{
cep 22290-180, Rio de Janeiro (RJ), Brazil.}}
~\\

\begin{abstract}

~\par
{\it Color Lie (super)algebras} were introduced in 1978 by Rittenberg and Wyler  as generalizations of ordinary Lie (super)algebras.
These color structures, graded by abelian groups, are defined by brackets which satisfy generalized Jacobi identities.
Only recently the relevance of color Lie (super)algebras to physical applications has been pointed out and proved. The color framework not only describes, but also provides physical signatures for two broad types of parastatistics (particles beyond ordinary bosons/fermions): \\
{\it i}) paraparticles which are exchanged via the permutation group and can exist in any space dimension $D$ and\\ {\it ii})  anyons (exchanged via the braid group), which exist up to $D=2$ space dimensions. \\
This paper presents an update of the recent advancements which motivate
a brand new ``colored look" at questions involving parastatistics and braid statistics.
~\par
\end{abstract}
\vfill
~
\newpage

\section{Introduction}
This paper gives a succinct introduction to ``{\it color Lie (super)algebras}", pointing out newly found motivations to apply them to physics. These mathematical structures (introduced in 1978 by Rittenberg-Wyler \cite{{rw1},{rw2}} as generalizations of ordinary Lie (super)algebras) are graded by abelian groups and defined by brackets which satisfy generalized Jacobi identities; their applications to physical systems imply two broad types of parastatistics:\\
{\it i}) paraparticles which are exchanged via the permutation group and can exist in any space dimension $D$ and\\ {\it ii}) anyons, which are exchanged via the braid group and exist up to $D=2$ space dimensions. \\
The color Lie (super)algebras, systematized by Scheunert in \cite{sch}, are a well-defined branch of mathematics recognized as 17B75 entry in the Mathematics Subject Classification MSC2010. Despite of that, the applications 
of the 17B75 color framework to physics  have been for many years rather sparse (they gained tractions in the last ten years only). A well-known reason was the misconception (going under the name of ``conventionality of parastatistics" or ``equivalence thesis", see \cite{conventionality} for an account) stating that paraparticles exchanged via the permutation group cannot be detected, their physical measurements being always reproducible by ordinary bosons/fermions. The two 2021 papers \cite{{top1},{top2}} proved for the first time that this is not the case; the hypotheses at the basis of the ``equivalence thesis" can be circumvented and physical signatures of parastatistics are encountered. The role of color Lie (super)algebras was crucial in providing a combinatorial counterexample to the equivalence thesis. Indeed, the \cite{top1} paper proved the detectability of
${\mathbb Z}_2\times{\mathbb Z}_2$-graded color parafermions, while \cite{top2} proved the same for ${\mathbb Z}_2\times{\mathbb Z}_2$-graded color parabosons.\\
More inexplicable is the complete absence of any color Lie (super)algebra investigation in the context of braid statistics.  The anyons (the quasi-particles theoretically predicted in \cite{lemy} and so named by Wylczek in \cite{wil}) were never controversial and recognized, e.g., for the role played in the fractional Quantum Hall Effect. Even so, it
turns out that the recent paper \cite{brqm}, published in May 2026, is the first one in defining concrete models, within a color Lie (super)algebra framework, which present physical signatures of braid statistics. \\
The present work provides a systematic quick introduction to the above topics stressing the relevant features and the newly discovered potentialities of color Lie (super)algebras. It is structured as follows:\\
Section {\bf 2} gives an updated state-of-the-art about the two types of parastatistics; Section {\bf 3} recalls the definition of {\it color Lie (super)algebras} following Scheunert's paper \cite{sch};  Section {\bf 4}  details the role of the ${\mathbb Z}_2^n$ grading ($n$-bit parastatistics) in the detectability of ``group-theoretical paraparticles"; Section {\bf 5} discusses the recent applications to braid statistics obtained from more general abelian grading groups (color Lie (super)algebras beyond $n$-bit). The Conclusions present comments about the above constructions and discuss future perspectives.

\section{Two types of parastatistics}
This Section outlines the differences between two types of parastatistics: the first one is defined by the permutation group, while the second one (anyonic parastatistics) is defined by the braid group. Both types of parastatistics admit two classes of paraparticles,
parabosons and parafermions, with parafermions satisfying a (generalized) Pauli exclusion principle. 
\subsection{The ``permutation-group parastatistics"}
The first attempt to introduce paraparticles beyond bosons/fermions was due to Gentile \cite{gen} who presented a generalization of the Pauli exclusion principle. In 1953 Green defined \cite{gre} a consistent framework, based on the so-called ``trilinear relations", to introduce either parabosonic {\it or} parafermionic oscillators. Later, this scheme was extended \cite{grme} to accommodate {\it both} parabosonic and parafermionic oscillators. Soon after their introduction, since paraparticles were not found in Nature, ‌physicists started to investigate possible explanations for their absence. A consensus emerged, the recalled ``equivalence thesis" of the \cite{conventionality} historical account: paraparticles do not produce distinct signatures of their presence, all their experimental results being reproduced and reinterpreted by ordinary bosons/fermions. The validity of this thesis is conditioned by the acceptance (and applicability) of certain hypotheses, with a {\it locality principle} presented in \cite{{drharo},{dhr}} playing a major role.\\
The \cite{top1} paper was the first one presenting an actual mathematical test of the equivalence thesis; a counterexample was found: parafermions are theoretically detectable; it was soon followed by \cite{top2}, which extended the counterexample to parabosons. Two extra papers with further counterexamples are \cite{nbits} and \cite{waha}. The \cite{{top1},{top2},{nbits}} tests discussed First-Quantized indistinguishable multiparticles, their quantum Hamiltonians being presented in the \cite{maj} Majid's  graded Hopf algebras framework; the \cite{waha} tests were applied to spin chains. In all these four works the locality's hypothesis at the core of the equivalence thesis was (via different mechanisms) evaded. Thanks to these works a new consensus (which received attention also in popularized press, see e.g. \cite{newsc}) emerged: under certain circumstances paraparticles can have observable consequences. 
\\
What about the experimental detection of ``permutation-group paraparticles"? They have not yet experimentally detected. An experimental group proved \cite{paraexp} how to manipulate parastatistics in the laboratory by using the spin of a trapped atomic ion and laser-inducing couplings between two of its excitation modes.  This result cannot yet be qualified as an experimental detection; the reason is the lack of a clear signature which guarantees that the outcome is not reproduced by ordinary bosons/fermions. Still, the importance of \cite{paraexp} is the demonstration that the manipulation of parastatistics is in the realm of feasibility.\\
Two recent preprints \cite{{gedanken},{wh2}} discuss how to apply the platonic mathematical tests into a doable experimental detection of paraparticles. The \cite{gedanken} proposal is based on encoding the required minimal logical operations into an experimental setting realized via quantum information technology.\\
The quest towards the detection of ``permutation-group paraparticles" is now an open challenge.
\subsection{The braid statistics (anyons)}

The existence of a new type of parastatistics in low space dimensions (up to $D=2$) was first pointed out by Leinaas and Myrheim \cite{lemy} in 1977. The corresponding paraparticles are exchanged via the  braid group; the name {\it anyons} was coined for them by Wilczek in 1982 \cite{wil}. A historical account of the theoretical discovery of anyons, with the relevant bibliography, is given in \cite{gol}.\\
We recall that the $n$-strand braid group $B_n$ is presented by $n-1$ generators $b_i$ ($i=1,\ldots, n-1$) satisfying
\bea
b_ib_{i+1}b_i =b_{i+1}b_ib_{i+1}\quad &{\textrm{and}}&\quad b_i b_j =b_jb_i {\textrm{\qquad for \quad $\vert i-j\vert\geq 2$.}} 
\eea
The $n$-element permutation group $S_n$ is recovered, as a special case, under the condition $b_i^2=1$.\\
It took 43 years, since their theoretical prediction, for anyonic quasi-particles to be experimentally detected in a laboratory.
The first experimental detection took place in 2020, see \cite{expanyons}. Later on, other experiments proving their existence followed \cite{{nonabanyons},{1danyons}}. \\
In recent years investigations on anyons occupied a central role for possible technological applications. This research program followed the Kitaev's proposal \cite{kit} that emergent Majorana quasi-particles in condensed matter could be used to encode quantum computations which offer topological protection to the decoherence suffered by ``ordinary quantum computers".
The ``knot logic" behind topological quantum computations with Majorana fermions was discussed, e.g., in \cite{kau}.\\
The 2025 Microsoft's announcement  \cite{microsoft} of the first quantum chip powered by a topological architecture opened the possibility,  see \cite{roadmap}, to a roadmap for the implementation of a fault tolerant quantum computation. The Kitaev's theoretical proposal is getting closed to be implemented as a feasible technological possibility. The anyonic braid statistics plays a central role in this technological challenge.
\section{Quick presentation of color Lie (super)algebras}

We recall for completeness the basic definitions of the \cite{{rw1},{rw2}} Rittenberg-Wyler's {\it color Lie (super)algebras}. It is more convenient, in many respects, to follow the Scheunert's presentation \cite{sch} (the Rittenberg-Wyler's conventions will be restored when needed).\\
It is worth mentioning, out of curiosity, that the term {\it color} was coined for a (failed) attempt  to apply these structures to quarks.\\
One starts with an abelian group $\Gamma$ and a
$\Gamma$-graded vector space $V=\oplus_\gamma V_\gamma$, $\gamma\in \Gamma$. The notion of a {\it commutation factor $\varepsilon$} on an abelian group $\Gamma$ is introduced. 
It is a bilinear map $\varepsilon: \Gamma\times \Gamma \rightarrow C^\ast$ (taking values in the punctured complex plane
${\mathbb C}^\ast\equiv {\mathbb C}\backslash \{0\}$) satisfying, for any $\alpha,\beta,\gamma$ in $\Gamma$, the properties:
\bea\label{epsprop}
&&~~{\textrm{{\it i}): \quad $\varepsilon(\alpha,\beta)\cdot \varepsilon(\beta,\alpha) =1$,}}\nonumber\\
&&~{\textrm{{\it ii}): \quad $\varepsilon(\alpha,\beta+\gamma) = \varepsilon(\alpha,\beta)\cdot \varepsilon(\alpha,\gamma)$,}}\nonumber\\
&&{\textrm{{\it iii}): \quad $\varepsilon(\alpha+\beta,\gamma) =\varepsilon(\alpha,\gamma)\cdot\varepsilon(\beta,\gamma)$.}}
\eea
A $\Gamma$-graded color Lie (super)algebra $L=\oplus_\gamma L_\gamma$  is defined by the brackets $\langle .,.\rangle: L\times L\rightarrow L$ which satisfy the following conditions, for any $A,B,C\in L$ and respective $\alpha, \beta, \gamma$ gradings:
\bea\label{epsilonalgebras}
&&{\textrm{{\it i}) $\quad \langle A, B\rangle = -\varepsilon (\alpha,\beta)\langle B, A\rangle\qquad \quad\quad  $ (the ``$\varepsilon$-skew" symmetry) and}}\nonumber\\
&&{\textrm{{\it ii})  $\quad \varepsilon(\gamma,\alpha)\langle A,\langle B,C\rangle\rangle + cyclic =0\quad $ (the ``$\varepsilon$-Jacobi" identities).}}
\eea
The first relation in (\ref{epsprop}) necessarily implies $\varepsilon(\alpha,\alpha) =\pm1$. \\
Due to physical motivations, the term {\it color Lie algebra} is employed if the sign $+1$ is valid for any $\alpha\in \Gamma$; otherwise, if there exists at least one $\alpha \in\Gamma$ such that
$\varepsilon(\alpha,\alpha) =-1$, the corresponding algebra is called a {\it color Lie superalgebra}.\\
For any associative $\Gamma$-graded space $S$ the bracket $\langle.,.\rangle$ defined by
\bea\label{gradedbracket}
\langle A, B\rangle := AB-\varepsilon(\alpha,\beta) BA
\eea
induces on $S$ a color Lie (super)algebra satisfying the equations (\ref{epsilonalgebras}).
Some remarks are in order:\\
{\it Remark 1}) the graded bracket defined at the special value $\varepsilon(\alpha,\beta)=+1$ 
corresponds to an ordinary commutator ($\langle A, B\rangle = AB-BA=[A,B]$) while,
for $\varepsilon(\alpha,\beta)=-1$, the graded bracket corresponds to an anticommutator
($\langle A, B\rangle = AB+BA=\{A,B\}$).\\
{\it Remark 2}) let us now specialize the analysis to the case of $\Gamma$ being a finite abelian group. In this case $\Gamma$ is expressed by a product of cyclic groups. The generic finite abelian $\Gamma$ is given by
\bea\label{productof}
\Gamma&=& \Pi_{i=1}^r {\mathbb Z}_{{p}^{k_i}_i},\quad {\textrm{with $r,k_i\in {\mathbb N}$ and the $p_i$'s prime numbers.}}
\eea
A subclass of the finite abelian groups is given by $p_i=2,~k_i=1,~r=n$:
\bea\label{z2n}
\Gamma &=& {\mathbb Z}_2^n\quad {\textrm{(which imply, as seen below, a $n$-bit parastatistics).}}
\eea
In this subclass the $\varepsilon(\alpha,\beta)$ commutation factors only take values $\pm 1$. Therefore,
their induced graded brackets (\ref{gradedbracket}), discussed in Section {\bf 4},  are only given by commutators or anticommutators. \\
The situation can differ for more general finite abelian groups (belonging to the ${\mathbb Z}_2^n$ subclass is a sufficient, but not necessary condition to satisfy the $\pm 1$ restriction on commutation factors).  The simplest example of a grading producing graded brackets (\ref{gradedbracket}) which interpolate commutators and anticommutators is given by $\Gamma = {\mathbb Z}_3\times {\mathbb Z}_3$. It will be discussed in Section {\bf 5}.

\section{The ${\mathbb Z}_2^n$-graded ($n$-bit) color Lie (super)algebras}

The term ``$n$-bit parastatistics" to denote the (\ref{z2n}) grading of the color Lie (super)algebras was introduced in \cite{nbits}.
For this subclass of abelian groups it is convenient to express the $\pm 1$ signs of the commutation factors via the angle position $\varepsilon(\alpha,\beta) = e^{i\pi\vartheta_{\alpha\beta}}$, so that $+1$ (the commutator) is mapped into $\vartheta_{\alpha\beta}=0$ and $-1$ (the anticommutator) is mapped to $\vartheta_{\alpha\beta}=1$.\\
It was shown in \cite{nbits} that the total number $b_n$ of inequivalent, consistent choices of  ${\mathbb Z}_2^n$-graded color Lie (super)algebras is given, in terms of the floor function,  by
\bea\label{floor}
b_n&=& n+\lfloor n/2\rfloor+1.
\eea
It produces, for $n=1,2,3,\ldots$, the $b_n$ series $2,4,5,7,8,10,\ldots$. \\
The two consistent $n=1$ choices produce an ordinary ${\mathbb Z}_2$-graded Lie algebra defined by commutators and an ordinary
${\mathbb Z}_2$-graded Lie superalgebra. \\
The first nontrivial color Lie (super)algebras are encountered at $n=2$ (``nontriviality" means that they are not reduced to ordinary Lie algebras or superalgebras). Out of the $b_2=4$ inequivalent  cases presented in the tables below, the nontrivial 
ones are respectively labeled CLA and CLSA: 
\bea \label{2cases}
{\textrm{LA:}}\qquad &&\begin{array}{c|cccc}&00&10&01&11\\ \hline 00&0&0&0&0\\10&0&0&0&0\\01&0&0&0&0\\11&0&0&0&0
\end{array} \qquad {\textrm{(an ordinary Lie algebra),}}
\nonumber\\
&&\nonumber\\
{\textrm{LSA:}}\quad&& 
\begin{array}{c|cccc}&00&10&01&11\\ \hline 00&0&0&0&0\\10&0&1&1&0\\01&0&1&1&0\\11&0&0&0&0
\end{array}\qquad
 {\textrm{(an ordinary Lie superalgebra),}}
\nonumber\\&&\nonumber\\
{\textrm{CLA:}}\quad&& 
\begin{array}{c|cccc}&00&10&01&11\\ \hline 00&0&0&0&0\\10&0&0&1&1\\01&0&1&0&1\\11&0&1&1&0
\end{array}\quad
 {\textrm{(a nontrivial color Lie algebra),}}\nonumber\\
&&\nonumber\\
{\textrm{CLSA:}}\quad&& 
\begin{array}{c|cccc}&00&10&01&11\\ \hline 00&0&0&0&0\\10&0&1&0&1\\01&0&0&1&1\\11&0&1&1&0
\end{array}\quad
 {\textrm{(a nontrivial color Lie superalgebra);}}\nonumber\\&&
\eea
The entries define the commutators $(0)$ or anticommutators $(1)$ associated with the respective pairs of $2$-bit gradings
($00,01,10,11$) of the color Lie (super)algebra generators. In physical applications, the $1$ diagonal entries are associated with fermionic-like particles (such as the $10$-graded and $01$-graded sectors of CLSA) which anticommute with themselves and satisfy the Pauli exclusion principle. One should note that, in CLSA, the $10$ and $01$ graded sectors correspond to parafermions which mutually commute instead of anticommuting. The CLA table presents three parabosonic sectors ($10$,
$ 01$,$11$) which mutually anticommute.\\
Some relevant remarks are:\\
{\it Remark 1}) the above $4\times 4$ CLSA table is a generalization of an ordinary Lie superalgebra, which is recovered by setting all bosons in the $00$ sector, all fermions in, let's say, the $10$ sector, while leaving empty the $01$ and $11$ sectors. \\
{\it Remark 2}) Comparing CLSA with LSA allows to spot the difference between a pair of oscillators accommodated in their respective $10$ and $01$ sectors  (the parafermionic ones in CLSA, the fermionic ones in LSA). In particular, the parafermionic (fermionic) creation operators $p_{10}^\dagger, p_{01}^\dagger$ ($f_{10}^\dagger, f_{01}^\dagger)$ respectively satisfy 
\bea
\{p_{10}^\dagger,p_{10}^\dagger\}=\{p_{01}^\dagger, p_{01}^\dagger\}=[p_{10}^\dagger,p_{01}^\dagger]&=&0\qquad {\textrm {and}}\nonumber\\
\{f_{10}^\dagger,f_{10}^\dagger\}=\{f_{01}^\dagger, f_{01}^\dagger\}=\{f_{10}^\dagger,f_{01}^\dagger\}&=&0,
\eea
so that
\bea\label{paraandfer}
p_{01}^\dagger p_{10}^\dagger = + p_{10}^\dagger p_{01}^\dagger &{\textrm{while}}& 
f_{01}^\dagger f_{10}^\dagger = -f_{10}^\dagger f_{01}^\dagger.
\eea
In both cases, since the difference is just a sign, exchanging twice 
\bea S_{12}: p_{01}^\dagger\leftrightarrow p_{10}^\dagger, && 
S_{12}: f_{01}^\dagger\leftrightarrow f_{10}^\dagger 
\eea
produces $S_{12}^2=1$; the closure of the permutation group is a feature of the ${\mathbb Z}_2^n$ grading.\\
The $\pm 1$ sign difference in the respective r.h.s. of (\ref{paraandfer})  turns out to have a major consequence which is exploited in the \cite{top1} minimal test.
\subsection{A minimal test to detect parafermions}
Bruce-Duplij introduced in \cite{brdu} a quantum mechanical model which is both invariant under a ${\mathbb Z}_2^2$-graded color Lie superalgebra and an example of supersymmetric quantum mechanics.  No physical signature from the ${\mathbb Z}_2^2$ grading is present in the original \cite{brdu} model. The situation changes in its First-Quantized, multi-particle version discussed in \cite{top1}. Then, the original creation operators are replaced by the respective coproducts: $\Delta(p_{10}^\dagger)$, $\Delta(p_{01}^\dagger)$ in the color Lie superalgebra case and $\Delta(f_{10}^\dagger)$, $\Delta(f_{01}^\dagger)$ in the Lie superalgebra case.  
It turns out \cite{{top1},{gedanken}} that, in the $2$-particle sector, an energy eigenstate is mapped into a $2$-component vector
{\footnotesize{$\left(\begin{array}{c} 1\\ \varepsilon\end{array} \right)$}}, with $\varepsilon =+1$ for parafermions and $\varepsilon =-1$ for ordinary fermions. Alice sends to Bob the suitably prepared eigenstate and Bob performs a yes-no projection measurement which determines the $\varepsilon=\pm 1$ outcome:
{\footnotesize{\bea\label{proj}
\frac{1}{2}\left(\begin{array}{cc} 1& 1\\ 1&1\end{array}\right)\left(\begin{array}{c} 1\\ \varepsilon\end{array} \right)&=&\frac{1}{2}(1+\varepsilon)\left(\begin{array}{c} 1\\ \varepsilon\end{array} \right).
\eea}}
In a series of repeated tests, if Bob's outcomes are different, he can conclude that particles of both $\pm 1$ parity (i.e. fermions and parafermions) are encountered.\\
More elaborate scenarios can produce more spectacular results; as an example, Table (7.31) of \cite{nbits} discriminates parastatistics from ordinary statistics by counting the degeneracy of certain energy levels. The advantage of the minimal scenario consists in the fact that, due to its simplicity, it looks promising to be physically implemented. In this respect, 
the $n$-bit presentation of the ${\mathbb Z}_2^n$-graded color Lie (super)algebras has a potentially useful feature. It allows to express the tables of the respective commutation factors as Boolean logic gates (see the Appendix {\bf B} of \cite{nbits}). Hopefully, this property could be exploited, as suggested in \cite{gedanken}, in encoding the abstract mathematical framework of the $n$-bit parastatistics into a physically doable experiment to detect engineered paraparticles.

\section{Color Lie (super)algebras beyond $n$-bit: braid statistics}

The $0,1$ ``angle notation" is useful to connect the ${\mathbb Z}_2^n$ grading to $n$ bits. For more general abelian groups 
$\Gamma$ it is more useful to revert to the (\ref{epsprop}) convention for the ${\varepsilon(\alpha,\beta)}$ commutation factors. Due to the fact that a generic $\Gamma$ is a product of cyclic groups, the ${\varepsilon(\alpha,\beta)}$ entries in the tables below are given by roots of unity.\\
For a generic grading abelian group $\Gamma$  the graded brackets (\ref{gradedbracket}) can result into a nontrivial linear combination of commutators/anticommutators. Indeed,
{\small{\bea\label{interpolation}
&\langle A, B\rangle = AB-\varepsilon(\alpha,\beta) BA=  \frac{1}{2}\left(1+\varepsilon(\alpha,\beta) \right)\cdot [A,B] + \frac{1}{2}\left(1-\varepsilon(\alpha,\beta) \right)\cdot \{A, B\}.&\nonumber\\&&
\eea}}
The analysis of the inequivalent solutions of the $\varepsilon(\alpha,\beta)$ commutation factors easily repeats the same steps  done for the ${\mathbb Z}_2^n$ grading. \\
It turns out that the ${\mathbb Z}_3$ group only produces a trivial solution which, for the ``trit" $ {\underline 0},{\underline 1},{\underline 2}$ notation of the ${\mathbb Z}_3$ generators, reads:
\bea\label{trivial}
&\begin{array}{|c|ccc|}\hline &{\underline 0}&{\underline 1}&{\underline 2}\\ \hline {\underline 0}& 1&1&1\\{\underline 1}&1&1&1\\ {\underline 2}&1&1&1 \\ \hline\end{array}
\eea
The above table represents an ordinary, ${\mathbb Z}_3$-graded, Lie algebra whose brackets are defined by commutators.\\ The minimal group presenting a nontrivial solution is  ${\mathbb Z}_3\times{\mathbb Z}_3$. Its two inequivalent choices for the commutation factors are expressed by the table\\
\bea
U=&\begin{array}{|c|ccc|ccc|ccc|}\hline &{\underline{00}}&{\underline{ 01}}&{\underline{02}}&{\underline{10}}&{\underline{11}}&{\underline{12}} &{\underline{20}}&{\underline{21}}&{\underline{ 22}} \\ \hline 
{\underline {00}}& 1&1&1&1&1&1&1&1&1\\
{\underline{ 01}}&1&1&1&j^2&j^2&j^2&j&j&j \\  
{\underline {02}}& 1&1&1&j&j&j&j^2&j^2&j^2\\ \hline
{\underline {10}}&1&j&j^2&1&j&j^2&1&j&j^2 \\ 
{\underline {11}}& 1&j&j^2&j^2&1&j&j&j^2&1\\
{\underline {12}}&1&j&j^2&j&j^2&1&j^2&1&j \\  \hline
{\underline {20}}& 1&j^2&j&1&j^2&j&1&j^2&j\\
{\underline {21}}&1&j^2&j&j^2&j&1&j&1&j^2 \\ 
{\underline {22}}& 1&j^2&j&j&1&j^2&j^2&j&1\\  \hline\end{array}&\label{z3z3array}
\eea
where $j$ is a third root of unity ($j^3=1$). The two inequivalent choices correspond to $j=1$ (an ordinary, ${\mathbb Z}_3\times{\mathbb Z}_3$-graded, Lie algebra) and $j$ given by  a {\it primitive} third root of unity (either $j=e^{\frac{2\pi i}{3}}$ or
$j=e^{\frac{4\pi i}{3}}$). The $j=e^{\frac{2\pi i}{3}}$ solution produces a nontrivial color Lie algebra whose graded sectors different from ${\underline {00}}$ are parabosonic (due to the $+1$, in Scheunert's notation, diagonal elements). In the $j=e^{\frac{2\pi i}{3}}$ colored case the ${\underline {00}}$ sector is the unique bosonic sector due to the string of $+1$ entries in its row/column (all brackets defined for ${\underline {00}}$-graded generators are ordinary commutators).\\
${\mathbb Z}_2\times{\mathbb Z}_3\times{\mathbb Z}_3$ is the simplest abelian grading group which accommodates colored parafermions whose nontrivial mixed brackets interpolate commutators and anticommutators. The extra 
${\mathbb Z}_2$ produces a $18\times 18$ table $U'$ which is schematically expressed as \bea\label{aiter}
U' &=& \begin{array}{|cc|}U&U\\U&\delta U\end{array}  \qquad {\textrm{for \quad $\delta=\pm 1$.}}
\eea
By setting $\delta = -1$ and, as before, $j=e^{\frac{2\pi i}{3}}$ one obtains colored parafermions.\\
Some remarks are in order:\\
{\it Remark 1}) the presence of commutation factors realized by roots of unity is not surprising due to the 
grading being realized by ${\mathbb Z}_{p^k}$ cyclic groups.\\
{\it Remark 2}) the ``mixed-bracket" parafermions satisfy, as shown below, a {\it generalized} Pauli exclusion principle, while the  ${\mathbb Z}_2^n$-graded parafermions satisfy the ordinary Pauli exclusion principle.

\subsection{Signature of mixed-bracket parabosons} 
The main question addressed in \cite{brqm} consisted in analyzing the physical consequences of the mixed-bracket color parastatistics by introducing quantum models of indistinguishable multi-particles. This required at first to introduce mixed-bracket generalizations of the Heisenberg-Lie algebras. We sketch here the basic strategy. At first ordinary graded matrices
$C_\alpha$ (in the case of the ${\mathbb Z}_3\times {\mathbb Z}_3$ grading they are nine $9\times 9$ matrices) are introduced. They are suitably chosen to be graded-commutative, so that \bea\label{gradedcomm}
\langle C_{\alpha},C_\beta\rangle := C_\alpha C_\beta -\varepsilon (\alpha,\beta)C_{\beta}C_{\alpha} &=&0.
\eea
With the help of the $C_\alpha$ graded matrices ordinary Heisenberg-Lie oscillators are mapped into mixed-bracket color Heisenberg-Lie oscillators. This construction is illustrated by considering the \cite{brqm} example for parabosonic oscillators (the construction of mixed-bracket parafermionic oscillators
proceeds along the same steps). \\
In the case of ${\mathbb Z}_3\times {\mathbb Z}_3$, up to $4$ bosonic oscillators 
$a_I, a_I^\dagger $ (for $I=1,2,3,4$) can be mapped into mixed-bracket Heisenberg-Lie oscillators $A_I, A_I^\dagger$ through the following positions. The bosonic Heisenberg-Lie algebra ${\mathfrak{h}}_{bos}(4)$ is defined as
\bea\label{4bososc}
&&{\mathfrak{h}}_{bos}(4)~ : \quad \{a_I,a_I^\dagger,c\}, \quad ~~~ {\textrm{where}}\nonumber\\
&&[a_I,a_J^\dagger ]=\delta_{IJ}\cdot c, \qquad [c,a_I]=[c, a_I^\dagger] =[a_I,a_J]=[a_I^\dagger,a_J^\dagger]= 0 \qquad \forall I,J.\nonumber\\&&
\eea 
 The parabosonic color Heisenberg-Lie algebra  ${\mathfrak{h}}_{pb}(4)=\{A_I,A_I^\dagger,C\}$ is introduced, in terms of 9 ${\mathbb Z}_3\times {\mathbb Z}_3$-graded commutative matrices $C_{\underline{ij}}$, via
\bea\label{genhpb4}
&& A_1 = C_{\underline{20}} \cdot a_1,\qquad\qquad A_1^\dagger = C_{\underline{10}}\cdot a_1^\dagger,\nonumber\\
&&A_2 = C_{\underline{22}} \cdot a_2, \qquad\qquad A_2^\dagger = C_{\underline{11}}\cdot a_2^\dagger, \qquad \nonumber\\
&& A_3 = C_{\underline{21}} \cdot a_3,\qquad\qquad A_3^\dagger = C_{\underline{12}}\cdot a_3^\dagger, \qquad\nonumber\\
&&  {{ A_4 = C_{\underline{02}}\cdot  a_4}},\qquad\qquad A_4^\dagger = C_{\underline{01}} \cdot a_4^\dagger, \qquad\nonumber\\
&&C~ = C_{\underline{00}}.
\eea
The grading of the l.h.s. generators coincides with the grading of the r.h.s. matrices $C_{\underline ij}$.\
The ${\mathfrak{h}}_{pb}(4)$ generators close a color mixed bracket Heisenberg-Lie algebra defined by
the relations {\small\bea\label{hpb4brackets}
\langle A_I,A_J^\dagger \rangle =\delta_{IJ}\cdot C, \qquad\langle C,A_I\rangle=\langle C, A_I^\dagger\rangle =\langle A_I,A_J\rangle=\langle A_I^\dagger,A_J^\dagger\rangle= 0 \quad && \forall I,J.\nonumber\\&&
\eea}
For the purpose of analyzing the physical signature of the mixed bracket parastatistics is sufficient to consider the subalgebra
spanned by a pair of creation/annihilation operators, let's say $I=1,2$. A two-particle oscillator Hamiltonian $H_{2p}$ can be defined as
\bea\label{2dimosc}
H_{2p}= \sum_{I=1}^2 A_I^\dagger A_I, &{\textrm{so that}}&
 [H_{2p}, A_I^\dagger] = A_I^\dagger.
\eea
The two-particle Fock Hilbert space ${\mathcal H}_{2p}$ is spanned by the energy eigenvectors 
$|n,m\rangle :=(A_1^\dagger)^n(A_2^\dagger)^m |vac\rangle$, where the creation operators $A_1^\dagger, A_2^\dagger$ are applied to the Fock vacuum$|vac\rangle$ satisfying 
$A_1|vac\rangle=A_2|vac\rangle=0.$ \\
The indistinguishability condition is implemented by requiring the indistinguishable Hilbert space ${\mathcal H}_{2p, ind}$ to be spanned by the energy eigenvectors
$(A_1^\dagger +A_2^\dagger)^n |vac\rangle$, where $n=0,1, 2,\ldots$.\\
It follows that the energy spectrum of the indistinguishable quantum model,  given by the $E_n=n$ non-negative integers, is not degenerate. \\
The signature of the color parastatistics appears, once prepared the quantum system in the $E_n=n$ energy eigenstate, in the physically measurable corresponding $p_n(x,y)$ probability density of finding the particles around the $x,y$ position. \\
Indeed, the probability density $p_n(x,y)$, for $n\geq 2$, is affected by the noncommutative properties of the $A_1^\dagger, A_2^\dagger$ colored creation operators. In the bosonic case the creation operators $a_1^\dagger, a_2^\dagger$ commute:
 \bea
 a_2^\dagger a_1^\dagger &=& a_1^\dagger a_2^\dagger.
 \eea
 This is no longer the case for the colored creation operators satisfying
 \bea
 &A_2^\dagger A_1^\dagger = z A_1^\dagger A_2^\dagger \neq A_1^\dagger A_2^\dagger \qquad {\textrm{for $z\neq 1$.}}&
 \eea
 In the above ${\mathbb Z}_3\times {\mathbb Z}_3$ example $z = j^2$, with $j$ a primitive third root of unity.\\
 The $S_{12}$ permutation exchange ($S_{12}^2=I$) defined by
 \bea
S_{12}:~ && A_1^\dagger \quad\leftrightarrow \quad A_2^\dagger  
\eea
induces, due to the noncommutativity,  a braid group representation $B_{12}$ 
acting on the polynomial $A_1^\dagger A_2^\dagger $. For the $z=j^2$ primitive third root of unity one gets
{\footnotesize{\bea
&B_{12}(A_1^\dagger A_2^\dagger) = A_2^\dagger A_1^\dagger = j^2\cdot A_1^\dagger A_2^\dagger,\quad  B_{12}^2(A_1^\dagger A_2^\dagger)= j\cdot A_1^\dagger A_2^\dagger, \quad B_{12}^3(A_1^\dagger A_2^\dagger)=  A_1^\dagger A_2^\dagger.&\nonumber \\&& 
\eea}} Therefore $B_{12}^3=I$, while $B_{12}^2\neq I$. This possibility requires an abelian grading group $\Gamma$ more general than ${\mathbb Z}_2^n$.\\
The noncommutativity of the creation operators imply a deformed Pascal triangle in the expansion
\bea
(A_1^\dagger+A_2^\dagger)^n&=&\sum_{k=0}^n  c_{n;k}(A_1^\dagger)^{n-k}(A_2^\dagger)^{k} {\textrm{\quad for \quad $k=0,1,\ldots, n$}}.
\eea
For $j$ a primitive third root of unity, the $c_{n;k} $ coefficients are read from the $k^{th}$ entry from the left in the $n^{th}$ row of the deformed Pascal triangle
\bea&
\begin{array}{cccccccc} 
n=0:\quad&1~~~&&&&&& \\
n=1:\quad&1~~~&1~~~&&&&&\\
n=2:\quad&1~~~&-j~~~&1~~~&&&&\\
n=3:\quad&1~~~&0~~~&0~~~&1~~~&&&\\
n=4:\quad&1~~~&1~~~&0~~~&1~~~&1~~~&&\\
n=5:\quad&1~~~&-j~~~&1~~~&1~~~&-j~~~&1~~~&\\
n=6:\quad& 1~~~&0~~~&0~~~&2~~~&0~~~&0~~~&1~~~\\
{\ldots}:&\ldots&&
\end{array} &
\eea
The above $c_{n;k}$ coefficients determine the measurable, normalized, $p_n(x,y)$ probability densities. At the second excited energy level $n=2$, the normalized parabosonic probability density $p^{pb}_2(x,y)$ (recovered from the eigenstate $\propto
|2,0\rangle -j|1,1\rangle+|0,2\rangle$) differs from the normalized bosonic probability density $p^{bos}_2(x,y)$ (recovered from the eigenstate $\propto
|2,0\rangle +2|1,1\rangle+|0,2\rangle$ determined by the undeformed Pascal triangle):
\bea
p^{pb}_2(x,y)&\neq&p^{bos}_2(x,y), {\textrm{\qquad as shown in  Fig. 2 and 3 in \cite{brqm}.}}
\eea


\subsection{Mixed-bracket parafermions and roots of unity truncations}
The above construction can be extended to mixed bracket color parafermionic oscillators; in this case, as shown in \cite{brqm}, it
produces the interesting phenomenon of the truncation, at roots of unity, of the indistinguishable multiparticle energy spectra. The essence of this phenomenon can be grasped by comparing the energy spectrum of three different sets ($s \equiv s_1,s_2,s_3$) of three ($I=1,2,3$) (para)fermionic  $P_{s;I}, {P_{s;I}}^\dagger$ oscillators. \\
The sets are respectively obtained from different abelian gradings:  \\
{\it i}) ${\mathbb Z}_2$ for $s_1$ (the corresponding oscillators are ordinary fermions), \\
{\it ii}) $ {\mathbb Z}_2^3\times  {\mathbb Z}_3^2$ for $s_2$ and \\
{\it iii}) $ {\mathbb Z}_2 \times  {\mathbb Z}_3^2$ for $s_3$. \\
The (para)fermionic nature of the creation operators ${P_{s;I}}^\dagger$ implies their nilpotency:
\bea\label{nil}
(P_{s;I}^\dagger)^2&=&0.
\eea
At a given $s$, the indistinguishable three-particle Hilbert space ${\mathcal H}_{s;ind}$  is spanned by the vectors 
\bea \label{3vec}
v_{s;n} := ({P_{s;1}}^\dagger+{P_{s;2}}^\dagger+{P_{s;3}}^\dagger)^n |vac\rangle &\in& {\mathcal H}_{s;ind},
 \eea
where $|vac\rangle$ is the Fock vacuum defined by $P_{s,I}|vac>=0$.\\
The corresponding $s$-labeled Hamiltonians (whose energy spectra $E_n$ are given by  non-negative integers) are
\bea
H_{s} := {P_{s;1}}^\dagger{P_{s;1}}+ {P_{s;2}}^\dagger{P_{s;2}}+{P_{s;3}}^\dagger{P_{s;3}}.
\eea
The nilpotency condition (\ref{nil})  implies that the non-vanishing vectors $v_{s;n}$ spanning ${\mathcal H}_{s;ind}$ are at most given by $n=0,1,2,3$.\\
The mixed brackets imply a non-anticommutativity of the parafermionic creation operators $P_{s,I}^\dagger$ expressed by
the relations 
{\footnotesize{\bea
&{P_{s;1}}^\dagger{P_{s;2}}^\dagger= a {P_{s;2}}^\dagger{P_{s;1}}^\dagger, ~
{P_{s;2}}^\dagger{P_{s;3}}^\dagger= b {P_{s;3}}^\dagger{P_{s;2}}^\dagger,~{P_{s;3}}^\dagger{P_{s;1}}^\dagger= c {P_{s;1}}^\dagger{P_{s;3}}^\dagger.\qquad&
\eea}}
Special values of the $a,b,c$ parameters, induced by different gradings, imply further constraints in the energy spectra:\\
{\it Case i}) The ${\mathbb Z}_2$ grading $s_1$ implies $a=b=c=-1$, producing $v_{s_1;2}={v_{s_1;3}}=0$ (i.e., the Pauli exclusion principle of ordinary fermions). Therefore,
\bea
{\textrm{the $H_{s_1}$ energy spectrum is:  $E=0,1$.}}&&
\eea
{\it Case ii}) The ${\mathbb Z}_2^3\times {\mathbb Z}_3^2$ grading $s_2$ implies $a=b= j$, $c=j^2$, where $j= e^{\frac{2}{3}\pi i}$ is a primitive third root of unity satisfying $1+j+j^2=0$. For these values one gets $v_{s_2;2}\neq 0$ and the special 
 ${v_{s_2;3}}=0$ truncation. Therefore,
\bea
{\textrm{the $H_{s_2}$ energy spectrum is:  $E=0,1,2$.}}&&
\eea
 {\it Case iii}) The ${\mathbb Z}_2\times {\mathbb Z}_3^2$ grading $s_3$ implies $a=b= -j$, $c=-j^2$ for $j= e^{\frac{2}{3}\pi i}$; $a,b,c$ are given by primitive {\it sixth} roots of unity. These values  imply that both $v_{s_3;2}\neq 0$ and
 ${v_{s_3;3}}\neq 0$. Therefore,
\bea
{\textrm{the $H_{s_3}$ energy spectrum is:  $E=0,1,2,3$.}}&&
\eea

\subsection{Braided Majorana qubits revisited in the color framework}

The mixed-bracket color Heisenberg-Lie (super)algebras provide a general framework to construct quantum models satisfying  braid statistics. For this scope one has to start with a suitably chosen finite grading abelian group $\Gamma$ which can accommodate both mixed-bracket parabosonic {\it and} parafermionic oscillators. This is a new framework which has not yet been properly investigated in the literature. As a special application, it was proved in \cite{brqm} that the multiparticle quantization of braided Majorana qubits can be recovered within this framework. Before sketching this color derivation, we briefly outline the
motivations for the introduction in \cite{topqubits} of ``braided Majorana qubits"; their braidings were realized via a braided tensor product. The \cite{topvoli} paper presented a second and a third scenario for the derivation of this model. The colored framework of \cite{brqm} corresponds to a fourth scenario which is the most suitable one for generalizations.\\
~\\
The main motivation of \cite{topqubits} was  to introduce a minimal setting for the topological quantum computation 
based on the Kitaev's proposal \cite{kit}. Let's remind that\\
1) ordinary computers manipulate bits as simplest basic elements;\\
2) similarly, ``ordinary" quantum computers manipulate qubits;\\
3) for topological quantum computation, a natural question would be: which could be the analogous, simplest basic elements? Following Kitaev, they should be Majorana fermions exchanged via the braid group. \\It was proposed in  \cite{topqubits} that the minimal Majorana fermion would correspond to the fermionic excited state of a two-state ${\mathbb Z}_2$-graded qubit.
 The vacuum state $|0\rangle$ is bosonic;  the fermionic excited state $ |1\rangle$ is created by the fermionic operator $\gamma$ acting on the vacuum:
 {\small{\bea\label{qubit01}
|0\rangle = \left(\begin{array}{c} 1\\0\end{array}\right) , &\quad& |1\rangle =\left(\begin{array}{c} 0\\1\end{array}\right), \qquad{\textrm{with ~~~$\gamma =\left(\begin{array}{cc} 0&0\\1&0\end{array}\right)$  .}}
\eea}}
The above system satisfies a reality condition; this means that the fermionic excited state coincides with its own antiparticle, i.e. it is a Majorana fermion.\par
Next, a representation of the braid group needs to be introduced in a multi-particle system of ${\mathbb Z}_2$-graded Majorana qubits. This is achieved following the Majid's prescription \cite{maj} of introducing a consistently defined ``braided tensor product"  which will be denoted with the symbol ``$\otimes_{br}$". For the creation operator $\gamma$ the braided tensor product satisfies
\bea \label{braidingamma}
({\mathbb I}_2\otimes_{br} \gamma)\cdot (\gamma\otimes_{br} {\mathbb I}_2) &=& B_t\cdot (\gamma\otimes_{br} {\mathbb I}_2)\cdot ({\mathbb I}_2\otimes_{br} \gamma) \equiv B_t\cdot (\gamma\otimes\gamma),\nonumber\\&&
\eea
where the $t$-dependent $4\times 4$ constant matrix $B_t$ is invertible for $t\neq 0$. $B_t$ coincides with the  $R$-matrix of the Alexander-Conway polynomial in the linear crystal rep on exterior algebra introduced in \cite{kasa}; it
is given by {\footnotesize{\bea\label{btmatrix}
B_t&=&\left(\begin{array}{cccc} 1&0&0&0\\0&1-t&t&0\\0&1&0&0\\0&0&0&-t\end{array}\right).
\eea
}}
 The consistency of the procedure is guaranteed by the fact that the matrix $B_t$ satisfies the braid relation
{\small{\bea\label{braidedrel}
(B_t\otimes {\mathbb I}_2)\cdot ({\mathbb I}_2\otimes B_t)\cdot 
(B_t\otimes {\mathbb I}_2) &=& ({\mathbb I}_2\otimes B_t) \cdot
(B_t\otimes {\mathbb I}_2)\cdot ({\mathbb I}_2\otimes B_t).
\eea}}
When $t$ is a root of unity ($t= t_k$, labeled by $k=2,3,4,\ldots$)  we have
\bea\label{ts}
(B_{t_k})^k&=&{\mathbb I}_4 \qquad {\textrm{ for \qquad $t_k = e^{\pi i (\frac{2}{k}-1)}$.}}
\eea
One constructs the $N$-particle Hamiltonians by repeatedly applying the Hopf algebra coproduct to the single-particle Hamiltonian {\footnotesize{$H =\left(\begin{array}{cc} 0&0\\0&1\end{array}\right)$}}; the multi-particle energy spectrum is derived
as a consequence. \par
The construction produces an implementation of a Gentile-type \cite{gen} parastatistics with a root-of-unity  truncation of the $N$-particle energy spectrum.
The energy levels $E$ are not degenerate and given by integer numbers. In function of $k$ one gets
\bea\label{truncatedenergy}
E &=& 0,1,\ldots, N\qquad \quad~\quad {\textrm{for}}\quad N<k,\nonumber\\
E &=& 0,1,\ldots, k-1 \qquad {\textrm{for}} \quad N\geq k;
\eea
a plateau is reached at the maximal energy level $k-1$, corresponding to the maximal number of braided Majorana fermions that can be accommodated in a multi-particle Hilbert space. \par
In the $k\rightarrow\infty$ limit $(t_\infty=-1)$ no plateau is reached and the maximal energy eigenvalues grow linearly with $N$:
\bea\label{genericenergy}
E &=& 0,1,\ldots, N\qquad {\textrm{for any given}}\quad N.
\eea
The following paper \cite{topvoli} further analyzed the model presenting two extra scenarios:
\\
{\it i}) it clarified that the level-$k$ truncations of the energy spectra are linked to roots-of-unity truncations from a quantum group; indeed,  the model can be recovered from superselected reps, at roots of unity, of the quantum superalgebra  ${\mathcal U}_q({\mathfrak{osp}}(1|2))$ introduced in \cite{kure} (see also \cite{{rbz1},{rbzlee}} for the special properties at roots of unity  of this quantum superalgebra);\\
{\it ii}) the braided tensor product $\otimes_{br}$ can be realized via the ordinary tensor product $\otimes$ provided that $2\times 2$ intertwining matrices $W_t$ are introduced. The mapping is expressed as follows:
\bea\label{matrixrepof}
 (\gamma~\otimes_{br}{\mathbb I}_2)\mapsto \gamma\otimes {\mathbb I}_2, &\qquad
 ({\mathbb I}_2~\otimes_{br}\gamma)\mapsto W_t\otimes \gamma,
\eea
where
\bea
 ({\mathbb I}_2~\otimes_{br}\gamma)\cdot 
 (\gamma~\otimes_{br}{\mathbb I}_2)&\mapsto& (W_t\otimes\gamma)\cdot(\gamma\otimes {\mathbb I}_2)= (W_t\gamma)\otimes\gamma,\nonumber\\
 (\gamma~\otimes_{br}{\mathbb I}_2)\cdot ({\mathbb I}_2~\otimes_{br}\gamma)&\mapsto&
(\gamma\otimes {\mathbb I}_2)\cdot (W_t \otimes\gamma) = (\gamma W_t)\otimes\gamma.
\eea
The $2\times 2$ 
intertwining operator $W_t$, whose explicit expression is given in \cite{topvoli}, satisfies the consistency condition
\bea\label{consistencyintertwining}
W_t\gamma &=& (-t) \gamma W_t.
\eea
The (\ref{matrixrepof}) explicit realization of the braided tensor product $\otimes_{br}$ in terms of ordinary matrices
allows to introduce parafermionic creation/annihilation operators $G_I^\dagger, ~G_I$. \\
The creation operators $G_I^\dagger$ satisfy
\bea\label{generalnilpotent}
({G}_I^\dagger)^2 &=& 0 {\textrm{\qquad\qquad \qquad (the nilpotency condition) for any $I$ and}}\nonumber\\
{G}_I^\dagger {G}_J^\dagger &=& r_k {G}_J^\dagger{G}_I^\dagger {\textrm{\qquad\quad  for $I<J$,\quad  with \quad $r_k=e^{\frac{2\pi i}{k}}$}}
\eea
($r_k$ is a primitive $k^{th}$ root of unity for $k=2,3,\ldots$).\\
The truncations of the multi-particle energy spectra is implied by the relations
{\footnotesize{\bea\label{combinatorics}
\left(\sum_{I=1}^{N} {G}_I^\dagger\right)^n \neq 0 {\textrm{\qquad for ~~~~$n<k$~~~~~~ and}} &~~~&
\left(\sum_{I=1}^{N} {G}_I^\dagger\right)^n = 0 {\textrm{\qquad for ~ $n\geq k$}}.
\eea}}
It was proved in  \cite{topvoli} that the $G_I, G_I^\dagger$ annihilation/creation operators close mixed-bracket
Heisenberg-Lie superalgebras defined by round brackets $(.,.)$. For any $X,Y$ pair of generators, suitable
$\vartheta_{x,y}$ angles are introduced such that
{\footnotesize{\bea\label{roundbracket}
\left(X,Y\right)_{\theta_{x,y}}&:= & i \sin (\theta_{x,y}) [X,Y]+ \cos (\theta_{x,y})\{X,Y\}~= ~e^{i\theta_{x,y}}\left(XY+e^{-2i\theta_{x,y}}YX\right).\nonumber\\
&&
\eea}}
It was shown in \cite{brqm} that the round brackets are reconstructed from the color Lie superalgebras $\langle.,.\rangle$  brackets (\ref{gradedbracket}) through the positions
\bea
\langle X, Y\rangle  &=& XY-\varepsilon(x,y)YX,\nonumber\\
\left(X, Y\right) &=& e^{i\theta_{x,y}} \langle X,Y\rangle\qquad {\textrm{for\quad $\varepsilon(x,y) = - e^{-2i\theta_{x,y}}$.}}
\eea

\section{Conclusions: comments and future perspectives}

This paper briefly discusses how color Lie (super)algebras can be used as a framework for physical applications of two broad classes of parastatistics:\\ {\it i})  in any space dimension $D$, paraparticles beyond bosons and fermions exchanged via the permutation group and, \\
{\it ii}) up to $D=2$ space dimensions, anyons exchanged via the braid group. \par
The first scenario, in any $D$, is implied by the grading abelian groups ${\mathbb Z}_2^n$ (the ``$n$-bit parastatistics); the second scenario for anyons requires, for grading, more general finite abelian groups $\Gamma$; in this general case $\Gamma$ is the product of ${\mathbb Z}_{p^k}$ cyclic groups, where $p$ is a prime number.\\
For many years physical applications of ${\mathbb Z}_2^n$-graded color Lie (super)algebras received limited attention due to the misconception (based on the ``conventionality argument") that their paraparticles cannot be directly observed. Mathematical investigations of colored parastatistics were conducted starting from \cite{{yaji1},{yaji2},{kaha},{kan},{tol2},{stvdj}}, but the physical applications lagged behind. The most notable exception is the recognition by Vasiliev in 1985 \cite{vas} that some of the problems of supergravity in de Sitter spacetime are cured by introducing a ${\mathbb Z}_2^2$-graded color structure. Vasiliev's paper appeared 10 years before the discovery of the accelerated expansion of the Universe; it gives nowadays a compelling 
argument advocating the role of ${\mathbb Z}_2^2$-graded parafermions in the context of Dark Energy.\\
In the last 10 years several works prompted a new boost in investigations on ${\mathbb Z}_2^n$-graded color physics.  It was recognized the importance of color Lie (super)algebras as symmetries of Partial Differential Equations \cite{aktt1} and systematic constructions of classical \cite{{akt1},{brusigma}} and quantum \cite{{brdu},{akt2},{que}} systems were conducted;  further investigations in many different directions followed. This is now a well-established and active research area. On top of that it is now clear under which conditions the hypotheses behind the conventionality argument can be evaded.\\
~\\
Rather strangely, physical applications of color Lie (super)algebras to braid statistics (anyons) are not existing 
(some physical developments, like the construction in \cite{{wito},{wito2}} of ${\mathbb Z}_4\times {\mathbb Z}_4$-graded color extensions of the Poincar\'e algebra were presented at a formal level since no specific model was investigated).
The \cite{brqm} paper is the first one devoted to this topic. A timeline can clarify part of the problem. Rittenberg-Wyler color Lie (super)algebras were introduced in 1978, while anyons started being investigated since the seminal 1982 Wilczek's paper \cite{wil}. In the meanwhile,
the mathematics of quantum groups started being developed in the decade of 1980's and the recognition of the special role
(truncations) of quantum groups at roots of unity appeared in the works of Lusztig and de Concini-Kac \cite{{lus},{dck}}.\\
The results of \cite{brqm} show that color mixed brackets, determined by roots of unity, are obtained from gradings induced by cyclic groups. This recognition points towards a new look at the old Rittenberg-Wyler's color Lie (super)algebra framework. A new class of physical models, presenting features similar to those of quantum groups at roots of unity, can be consistently formulated in the color approach; this requires a grading realized by a finite abelian group $\Gamma$ more general than ${\mathbb Z}_2^n$.
The color reconstruction of the \cite{{topqubits},{topvoli}} braided Majorana qubits is a very specific example of this general phenomenon.\\
It is worth pointing out that some mathematical structures introduced in physical applications can be profited in the color paradigm. This is the case of the ``ternary physics" investigated in \cite{kercqg} and based on a ${\mathbb Z}_3$ grading. As shown in table (\ref{trivial}) the ${\mathbb Z}_3$ grading is too poor to introduce a nontrivial color structure  (at least a ${\mathbb Z}_3\times {\mathbb Z}_3$ grading is required). Nevertheless, $3\times 3$ ${\mathbb Z}_3$-graded matrices, as those introduced in \cite{akl}, are used as building blocks to construct the ${\mathbb Z}_3\times{\mathbb Z}_3$ graded-commutative matrices  satisfying (\ref{gradedcomm});  the relation between ternary and color structures is investigated in \cite{caraf}.\\
As mentioned, the last ten years brought attention to the ${\mathbb Z}_2^n$-graded color physics. It seems that now the focus is shifting into enlarging the color paradigm to the general class of abelian group grading. The recent mathematical revisitation, with new results, of color Lie (super)algebras graded by an arbitrary abelian group \cite{rbzha}  is one of the indications of this very recent shifted 
attention.


\subsection*{Acknowledgment}

This work has been supported by the CNPq grant PQ 304747/2026-2.\\ 
I am especially indebted to Gerald G. Goldin for enlightening exchanges of ideas on the role of parastatistics and braid statistics. Special thanks are due to my collaborator Zhanna Kuznetsova who stressed the relevance of color Lie (super)algebras graded by general abelian groups.



\end{document}